\documentclass[10pt]{article}
\usepackage{graphicx}
\usepackage{subfigure}
\usepackage{amsmath}
\usepackage{amssymb}
\usepackage{multicol}
\usepackage{lipsum}
\usepackage{blindtext}

\makeatletter
\newenvironment{tablehere}
{\def\@captype{table}}
{}
\newenvironment{figurehere}
{\def\@captype{figure}}
{}
\makeatother

\begin{document}
\begin{center}
\begin{large}
\title\\{ \textbf{Effective String Theory Inspired Potential and Meson Masses in Higher Dimension.}}\\\
\end{large}

\author\

\textbf{$ Sabyasachi\;Roy^{\emph{a}}\footnotemark \:\:and\: D\:K\:Choudhury^{\emph{b,c}}$ } \\

\footnotetext{Corresponding author. e-mail :  \emph{sroy.phys@gmail.com}}
\textbf{a}. Centre for Theoretical Physics, Karimganj College, Karimganj-788710, India.\\
\textbf{b}. Department of Physics, Gauhati University, Guwahati-781014, India.\\
\textbf{c}. Physics Academy of the North East, Guwahati-781014, India. \\

\begin{abstract}
Nambu-Goto action in classical bosonic string model for hadrons predicts quark-antiquark potential to be\cite{Nambu-Goto} $V(r)=-\frac{\gamma}{r}+\sigma r +\mu_0$. In this report we present studies of masses of heavy flavour mesons in higher dimension with our recently developed wave functions obtained following string inspired potential. We report the dimensional dependence of the masses of mesons. Our results suggest that as the meson mass increases with the number of extra spatial dimension, it will attain the Planck scale ($ \sim 10^{19}GeV$) asymptotically at an astronomically large spatial dimension (we call it Planck dimension) $D_{Planck} \sim 10^{11}$, which sets the limit of applicability of Schrodinger equation in large dimension.
\end{abstract}
\end{center}
\

\begin{multicols}{2}
\section{Introduction:}\rm
Heavy flavour mesons with unequal-mass quarks have been a subject of extensive theoretical investigation because of their simplified dynamics \cite{Bjorken,Neubert}. Analysis of masses of mesons gives a better and comprehensive insight into the decay widths and other properties \cite{Vinodkumar1,Vinodkumar2}, which in turn opens up the way to study the CKM matrix elements \cite{Isgur1}. For such systems, non-relativistic potential model approach has been successful in making fairly acceptable predictions. \\
In this regard, a realistic potential model in higher dimension for quark-antiquark bound systems \cite{Bali,Bali2} is being predicted from Nambu-Goto action for free bosonic strings\cite{NG1,NG2,NG3}, which can be expressed as\cite{Nambu-Goto}:
\begin{equation}
 V(r)=-\frac{\gamma}{r}+\sigma r +\mu_{0}
\end{equation}
This is Cornell type\cite{Eicheten} linear plus Coulombic potential for mesons in higher dimension. Here, $ \frac{\gamma}{r}$ is the universal L\"{u}scher term with $\gamma = \frac{\pi(d-2)}{24}$ as the L\"{u}scher coefficient\cite{Luscher1}. $\sigma$ is the string tension, whose value is $ 0.178\; GeV^2$ \cite{Sigma}and $\mu_0$ is a scale parameter. $d$ is the space-time dimension with $d=D+1$, D being the spatial dimension. \\
Here it is worth mentioning that in this Cornell type potential,  linear term is basically dominant with Coulombic term appearing as the first order correction to it (eg, Arvis potential of reference \cite{Arvis}).  The basic difference with the Cornell potential is that, here the coefficient of $1/r$ term  (L\"{u}scher term) does not contain any gauge term and is only dimension dependent. This L\"{u}scher term being leading order correction term to linear term, such approximation is generally valid for short $r$. However, this can be made compatible for large $r$ by making proper choice of string tension $\sigma$ and tuning the coefficient the L\"{u}scher correction term. However, in our analysis, for simplification, we have worked with fixed value of $\sigma$ and increasing values of dimension parameter $D$ in $\gamma$ to explore the variation of meson masses. \\
With this potential based on effective string theory for flux tube, recently we have developed meson wave function in higher dimension, by solving Schrodinger wave equation in higher dimension\cite{Dong,Gang} following quantum mechanical perturbation technique for both the cases - (a) L\"{u}scher term as parent(unperturbed term) with linear term as perturbation  \cite{SR1,SR2} and (b) linear term as parent (unperturbed term) with L\"{u}scher term as perturbation\cite{SR3}. In this letter, we report ground state masses of heavy-light mesons in higher dimension making use of these wave functions. We study the dimensional dependence of masses of mesons and also compare our results in three dimension with recent theoretical and experimental expectations. As the meson masses increase with dimension, here we explore possibility of some higher cut-off limit of dimension which is compatible with the Planck mass limit.
\section{Formalism:}\rm
Pseudoscalar meson mass can be computed from the following relation \cite{Vinodkumar1,KKP}:
\begin{eqnarray}
M_P = m_Q +m_{\overline{Q}}+ \triangle E  \;
where,\;  \triangle E = <H>
\end{eqnarray}
In D-spatial dimension, the Hamiltonian operator $H$ has the form \cite{GRKhan}:
\begin{equation}
H=-\frac{\nabla_D^2}{2\mu}+V(r)
\end{equation}
Here,$\mu=\frac{m_Q m_{\overline{Q}}}{m_Q+ m_{\overline{Q}}}$ is the reduced mass of the meson with $m_Q$ and $m_{\overline{Q}}$ are the quark and antiquark masses; $V(r)$ is the inter-quark potential given in equation (1) and $\nabla_D^{2}$ is the Laplace's operator in D dimension which at $l=0$ is given by \cite{Laplace}:
\begin{equation}
\nabla_D^{2} \equiv \frac{d^{2}}{dr^{2}}+\frac{D-1}{r}\frac{d}{dr}
\end{equation}
Now, $<H>$ can be expressed as:
\begin{eqnarray}
<H> = <-\frac{\nabla_D^2}{2\mu}> + <-\frac{\gamma}{r}> + <\sigma r + \mu_0> \nonumber \\
= <H_1> + <H_2> + <H_3>
\end{eqnarray}
\subsection{Mass with L\"{u}scher term as parent:}\rm
The meson wave function within string inspired potential model considering L\"{u}scher ($-\frac{\gamma}{r}$) term as parent and linear term ($\sigma r$) as perturbation (with $\mu_0 =0$)has been reported in \cite{SR1,SR2} as:
\begin{equation}
\Psi(r,D)=N_1[1-K(D)r^{2}]r^{\frac{D-3}{2}}e^{-\mu\gamma r}
\end{equation}
Where, $K(D)=\frac{\sigma (2D-3)}{6\gamma}$,$C_D=\frac{\pi^{D/2}}{\Gamma(D/2+1)}$ \cite{Noura} and $N_1$ is the normalisation constant which is given as \cite{SR2}:
\begin{equation}\scriptsize
N_1=\frac{1}{(D.C_D)^{1/2}}.\frac{1}{[\frac{\Gamma(2D-3)}{(2\mu\gamma)^{2D-3}}-2.K.\frac{\Gamma(2D-1)}{(2\mu\gamma)^{2D-1}}+K^2.\frac{\Gamma(2D+1)}{(2\mu\gamma)^{2D+1}}]^{1/2}}
\end{equation}
Using this wave function, we compute terms in $<H>$ as (details discussed in Appendix:A):
\begin{eqnarray}
<H_1>=-\frac{D C_D N_1^2}{2\mu}\Sigma_{i=1}^7 J_i I_i \\
<H_2>=-\gamma D C_D N_1\Sigma_{i=1}^3 J_{7+i} I_i \\
<H_3>=D C_D N_1^2\Sigma_{i=1}^6 J_{10+i} I_{2+i}
\end{eqnarray}
The explicit expressions of integrals $I_i$s and parameters $J_i$s are given in equations (A.8-A.14,A.27) and (A.31) of Appendix:A.
\subsubsection{In the large $D$ limit:}\rm
We also extend our formalism for computing $<H>$ in the large $D$ limit. Here, we take:
\begin{equation}
DC_DN_1^2= \frac{1}{L_0}
\end{equation}
Where from eqn (7) we find:
\begin{equation}
L_0 = \frac{\Gamma(2D-3)}{(2\mu\gamma)^{2D-3}}-2K\frac{\Gamma(2D-1)}{(2\mu\gamma)^{2D-1}}+K^2\frac{\Gamma(2D+1)}{(2\mu\gamma)^{2D+1}}
\end{equation}
In the large $D$ limit, $D$ dependent terms transform as shown in equations (A.32 - A.41) of Appendix-A.
\scriptsize
\normalsize
Under such approximation, terms involved in $<H>$ now become:
\begin{eqnarray}\footnotesize
<H_1>  =-\frac{D^2}{L_c} \sum_{i=1}^7 L_i(\frac{2}{\pi\mu/12})^{i-1}  \\
<H_2>  =-\frac{D}{L_c}\sum_{i=1}^3 L_{7+i}(\frac{2}{\pi\mu/12})^{2i-1}\\
<H_3>  =\frac{1}{L_c}\sum_{i=1}^6 L_{10+i}(\frac{2}{\pi\mu/12})^{i+1}
\end{eqnarray}
Here, the specific $L_i$ terms, in large $D$ limit are D-independent and are given in equations (A.42-A.57) of Appendix-A. \\
It is thus found that, in the large $D$ limit, $<H>$ and hence masses increase as $D^2$, controlled mainly by the kinematic term $<H_1>$.
\subsubsection{Mass with only L\"{u}scher Term:}\rm
With only Luscher term in potential ($V(r)=-\frac{\gamma}{r}$, $<H> = <H_1> + <H_2>$), the wave function is:
\begin{equation}
\Psi^0(r)=Nr^{\frac{D-3}{2}}e^{-\mu\gamma r}
\end{equation}
Here, normalisation constant $N$ is:
\begin{equation}
N=\frac{1}{(D.C_D)^{1/2}}.\frac{(2\mu\gamma)^{2D-3}}{\Gamma(2D-3)}
\end{equation}
We ultimately find:
\begin{equation}\scriptsize
<H>=-\frac{DC_DN^2}{2\mu}[g_{11}I_1-g_{12}I_2 +g_{13}I_3]-\gamma DC_D N^2 I_2
\end{equation}
We have used:
\begin{eqnarray}
DC_DN^2=\frac{(2\mu\gamma)^{2D-3}}{\Gamma(2D-3)}  \\
g_{11}= g_1  ,\; g_{12}= g_2  \\
g_{13}= (g_3)_{K=0}=\mu^2 \gamma^2
\end{eqnarray}
In the large $D$ limit, equation (18) transform as:
\begin{equation}
<H>   = D^2\frac{1}{2\mu}(\frac{\mu\pi}{12})^2 \frac{1}{16} - D\mu (\frac{\pi}{24})^2
\end{equation}
Here also, thus, meson mass increases with $D^2$ as evident from equation (22). It might even reach the Planck scale $M_{Planck}\sim 10^{19} GeV$ at an astronomically large dimension $D$.

\subsection{Mass with linear term as parent:}\rm
With linear term as parent (and L\"{u}scher term as perturbation), the meson wave function has been calculated \cite{SR3} as:
\begin{eqnarray}
\Psi(r,D)= N_1r^{\frac{(1-D)}{2}}[1+A_1(r,D)r \nonumber \\
+A_2(r,D)r^{2}+.........](\varrho_1 r)^{m}Ai[\varrho_1 r-\varrho_0]
\end{eqnarray}
Here, terms involved have their explicit meaning as stated in ref. \cite{SR3}. The normalisation constant $N_1$ in $D$ spatial dimension can be calculated from the equation:
\begin{equation}
D C_D\int_0 ^\infty r^{D-1} |\Psi(r,D)|^2dr =1
\end{equation}
This expression for normalisation constant involves integration in $D$-fold space (D-spatial dimension). The surface element in $D$ spatial dimension is $DC_D r^{D-1}dr$, where the explicit expression of the terms $C_D$ has been stated in section 2.1.\\
As the wave function in this case contains Airy's infinite polynomial series, we calculate $<H>$ through numerical integration using the following integrals:
\begin{eqnarray}
<H_1>=\int_0^{\infty} D C_D r^{D-1}\Psi(r)[-\frac{\nabla_D^2}{2\mu}]\Psi(r)dr \\
<H_2>= \int_0^{\infty} D C_D r^{D-1}\Psi(r)[-\frac{\gamma}{r}]\Psi(r)dr \\
<H_3> = \int_0^{\infty} D C_D r^{D-1}\Psi(r)[\sigma r ]\Psi(r)dr
\end{eqnarray}
In this case, with Airy's function involved in the wave function, simpler equations like (13-22) in the large $D$ limit are not possible as we follow numerical integration to evaluate terms $<H_1>,<H_2>,<H_3>$.
\section{Results and conclusion:}\rm
With the different expressions for $<H>$, for two cases of Luscher parent and linear parent, we now proceed to calculate masses of $D, D_s, B, B_s$ and $B_c$ heavy flavoured mesons in higher dimension. We take the quark mass parameters from reference \cite{Vinodkumar2}. Here, we set the mass scale parameter $\mu_0$ by making our calculated meson masses in three dimension compatible with the corresponding standard PDG masses \cite{PDG}. Results are shown in Tables 1,2 and Figure 1,2.
\begin{tablehere}\scriptsize 
\begin{center}
\caption{Dimensional dependence of meson masses (in GeV) with L\"{u}scher term as parent}
\begin{tabular}{|c|c|c|c|c|c|}
  \hline
           & $D^0$    & $D_s$ & $B^0$   & $B_s$ & $B_c$  \\
  $\mu_0\rightarrow$  & 0.074    & 0.182  & 0.165 & 0.256 & 0.260    \\
    \hline  \hline
    D    &  $M_p$    &  $M_p$     &  $M_p$     &  $M_p$   &  $M_p$ \\
    \hline
    3    &  1.8649   &  1.9685    & 5.2796     & 5.3668   &  6.2770  \\
    5    &  1.8002   &  1.9201    & 5.2207     & 5.3235   &  6.2665  \\
    10   &  1.7645   &  1.8932    & 5.1881     & 5.2998   &  6.2615  \\
    20   &  1.7507   &  1.8840    & 5.1757     & 5.2916   &  6.2671  \\
    30   &  1.7486   &  1.8831    & 5.1751     & 5.2924   &  6.2832  \\
    40   &  1.7503   &  1.8876    & 5.1766     & 5.2968   &  6.3084  \\
    50   &  1.7543   &  1.8930    & 5.1812     & 5.3039   &  6.3425  \\
    100  &  1.7986   &  1.9533    & 5.2302     & 5.3725   &  6.6451  \\
    150  &  1.8786   &  2.0599    & 5.3179     & 5.4932   &  7.1671  \\
    200  &  1.9929   &  2.2118    & 5.4432     & 5.6650   &  7.9442  \\
    250  &  2.1413   &  2.4088    & 5.6058     & 5.8877   &  8.8684  \\
  \hline
\end{tabular}
\end{center}
\end{tablehere}

\begin{tablehere}\scriptsize
\begin{center}
\caption{Dimensional dependence of meson masses (in GeV) with linear term as parent}
\begin{tabular}{|c|c|c|c|c|c|}
  \hline
           & $D^0$    & $D_s$ & $B^0$   & $B_s$ & $B_c$  \\
  $\mu_0\rightarrow$  & 0.002    & 0.066  & 0.07 & 0.13 & 0.072    \\
    \hline  \hline
    D    &  $M_p$    &  $M_p$     &  $M_p$     &  $M_p$   &  $M_p$ \\
    \hline
    3    & 1.8649    & 1.9685 &  5.2796  &  5.3668  &  6.2770 \\
    5    & 1.9270    & 2.0093 &  5.3276  &  5.4031  &  6.2419 \\
    10   & 2.0569    & 2.1230 &  5.4521  &  5.5102  &  6.2796 \\
    20   & 2.3278    & 2.3668 &  5.7139  &  5.7433  &  6.4180 \\
    30   & 2.5755    & 2.5901 &  5.9534  &  5.9570  &  6.5489 \\
    40   & 2.8027    & 2.7944 &  6.1629  &  6.1523  &  6.6681 \\
    50   & 3.0141    & 2.9843 &  6.3771  &  6.3338  &  6.7781  \\
   100   & 3.9267    & 3.8021 &  7.2577  &  7.1142  &  7.2451 \\
   150   & 4.7014    & 4.4949 &  8.0049  &  7.7749  &  7.6372 \\
   200   & 5.3968    & 5.1163 &  8.6754  &  8.3673  &  7.9874 \\
   250   & 6.0381    & 5.6891 &  9.2936  &  8.9132  &  8.3101 \\
  \hline
\end{tabular}
\end{center}
\end{tablehere}

\begin{figurehere}
\centering
        {
        \includegraphics[width=3 in]{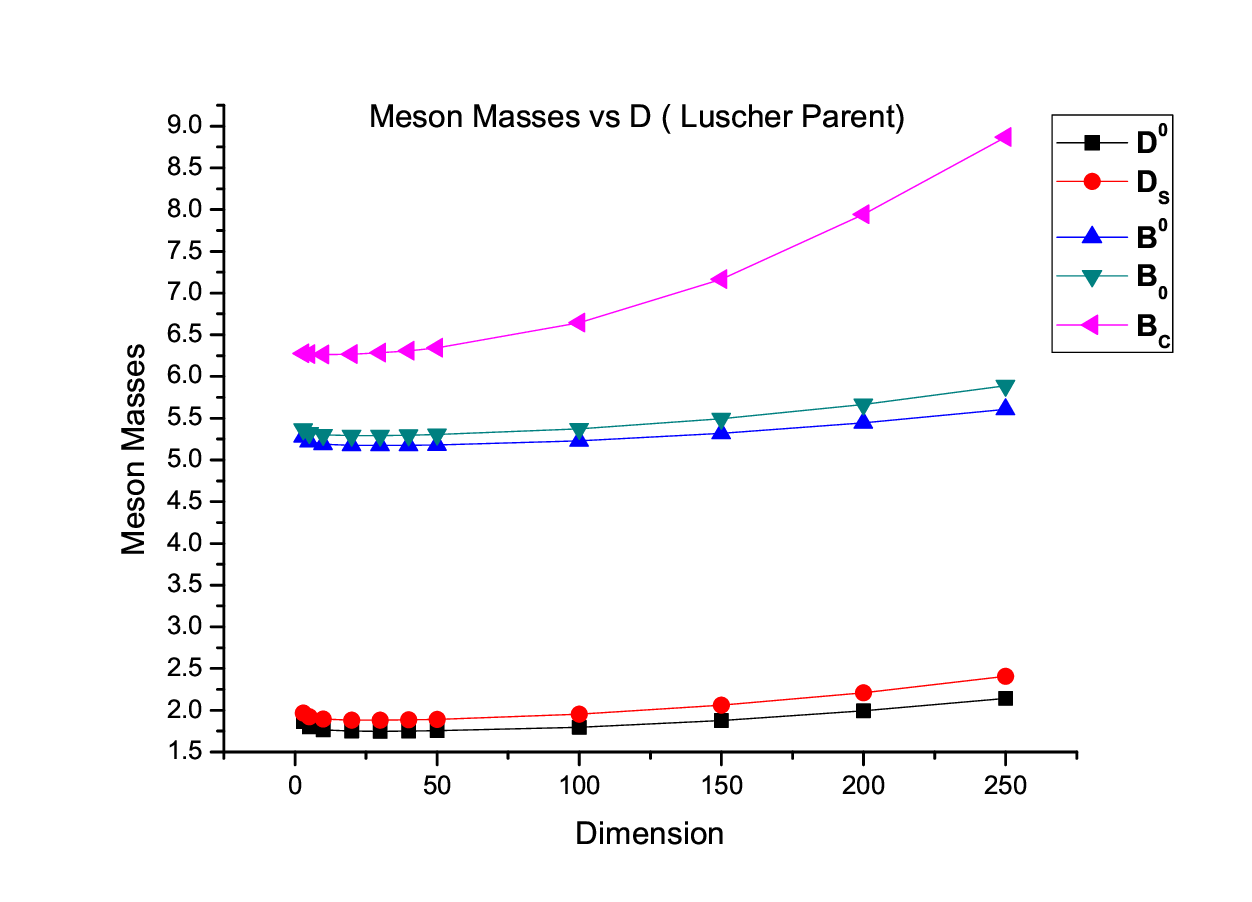}
        \label{fig:first_sub}
    }

    \caption{Meson mass vs dimension:Luscher parent}

\end{figurehere}

\begin{figurehere}
\centering
        {
        \includegraphics[width=3 in]{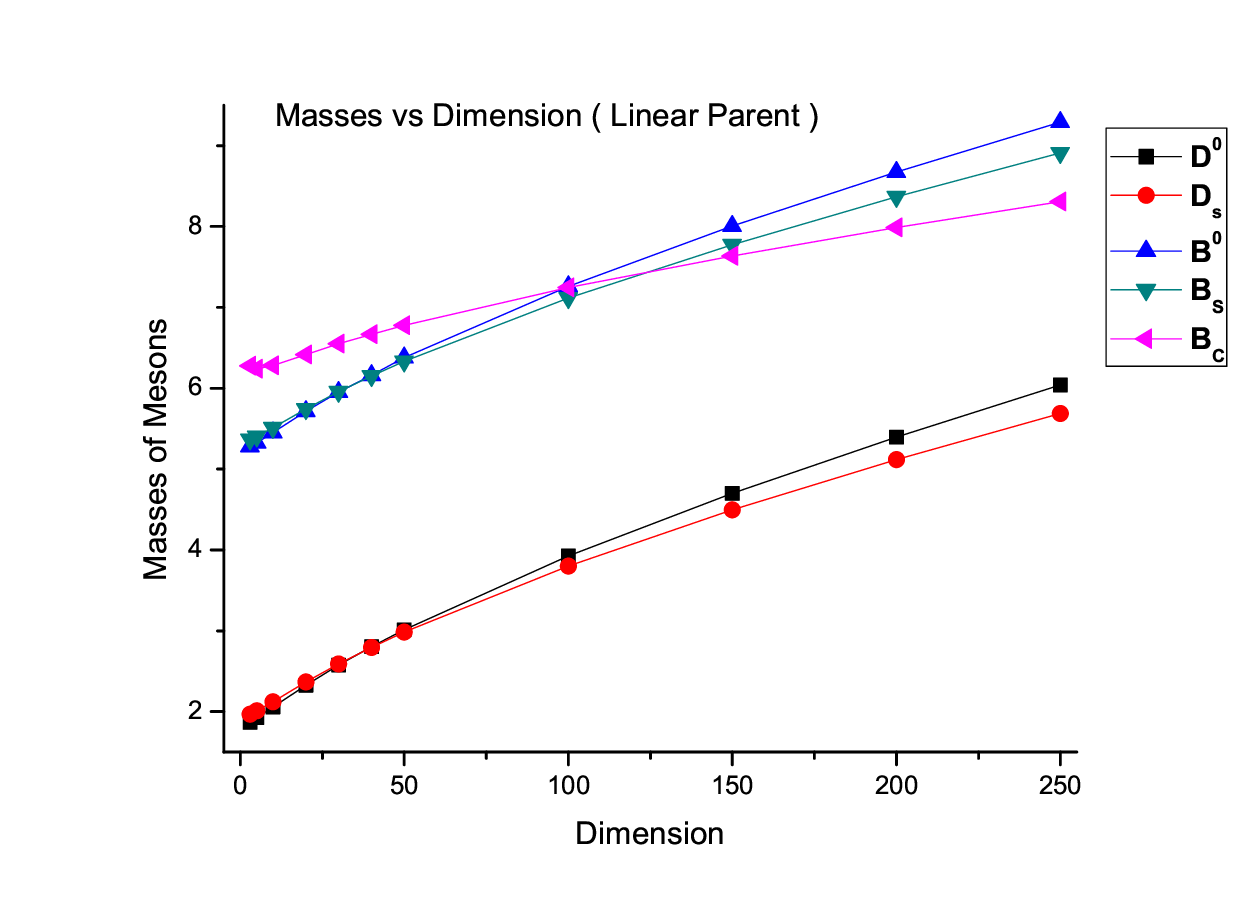}
        \label{fig:first_sub}
    }

    \caption{Meson mass vs dimension:linear parent}

\end{figurehere}

The above tables and figures show that, as the dimension increases masses also increase. Each of the terms in $<H>$ is also found to follow the same pattern in both the cases. However, the rate of increase of meson masses with dimension depends upon the flavour involved in the meson as well as on the choice of  `parent-child'. As an illustration, with linear parent, a typical D-meson mass becomes double at around $D=100$ (Table 3); however, for B meson such twofold increase in mass occurs at around $D=300$. \\
An interesting consequence of this pattern is that at a very large value of spatial dimension (may be called as Planck dimension $D_{Planck}$) the masses will attain the cut-off value of Planck mass $M_{Planck} \sim 10^{19} GeV$ above which the quantum mechanics of higher dimension cannot be applied. \\
Analytic expression for this asymptotic dimension is not possible in general, except in some special cases. \\
As an illustration, for $D^2 >> D $ limit, the equation (22) can be written as:
\begin{equation}
<H>   = D^2\frac{1}{32\mu}(\frac{\mu\pi}{12})^2
\end{equation}
This will yield the Planck dimension as:
\begin{equation}
D_{Planck} \approx \frac{48}{\pi}\sqrt{\frac{2M_{Planck}}{m_q}}
\end{equation}
in the infinite heavy quark mass limit ($\mu \rightarrow m_q $). Here, $M_{Planck}= \sqrt{\frac{\hbar c}{G}}\sim 1.22 \times 10^{19} GeV$ and $m_q$ is the mass of the light quark constituent. For typical $m_q=0.2GeV$, the numerical value of the Planck dimension is:
\begin{equation}
D_{Planck} \approx 1.6 \times 10^{11}\; (for \;\; m_q=0.2 GeV)
\end{equation}
This astronomically large value of dimension sets the natural limit of the applicability of the present quantum mechanical approach based on higher dimensional Schrodinger equation. \\
Further, we find through our analysis that the linear parent case is perturbatively stable while the other one - Luscher parent is not, within the model parameters used in the work.  We also observe that the kinematic term $<H_1>$ is the most dominant contributor in higher dimension. We presumably conclude that this may be a general feature of the Schrodinger equation in higher dimension and for any reasonable potential, the solution of Schrodinger equation might yield similar behavior of kinematic term in higher dimension. But, independent of perturbative stability or otherwise, both the options indicate that meson masses increase with dimension parameter.\\
Lastly, we conclude by making the following comment on the limitations of the model and the scopes for improvement. \\
(a)In our analysis, we have worked with fixed value of string tension $\sigma$. The effect of quark mass on string tension and the subsequent consequence on our analysis will be an interesting  analysis, which we plan to carry out in future. \\
(b)The present work is based on a specific string inspired potential model and does not incorporate asymptotic freedom and hence the running of the strong coupling constant  $\alpha_s(Q^2)$. The formalism can be extended to incorporate such effect in future by adopting higher dimensional Coulomb potential of QCD \cite{Morales,Burgbacher, Caruso} suitably.
(c)In this work, we have considered all the space dimensions to be large. In fact, the analysis will be more physical and convincing if the number of large dimensions is restricted to four and the extra dimensions are considered as compact ones. In that case, the work can be extended to derive physical interpretation on masses of mesons in LED which then can be applied to study other static and dynamic properties of mesons. This is a major scope for improvement of the present work which is under consideration.

\paragraph{Acknowledgement:}
\begin{flushleft}
\emph{SR acknowledges the support and valuable advices extended by Dr. Pushan Majumder of Indian Association of Cultivation of Science.}
\end{flushleft}

\end{multicols}

\appendix

\numberwithin{equation}{section}
\begin{center}
\section{Appendix:A}
\end{center}
\begin{center}
\textbf{\emph{Determination of integrals involved in $<H>$ in the case of L\"{u}scher term as parent }}
\end{center}
\begin{eqnarray}
<H_1> = \int_0^{\infty} D C_D r^{D-1}\Psi(r)[-\frac{\nabla_D^2}{2\mu}]\Psi(r)dr \\
= -\frac{D C_D}{2\mu}\int_0^{\infty} r^{D-1}\Psi(r)[\frac{d^2\Psi(r)}{dr^2}+\frac{D-1}{r}\frac{d\Psi(r)}{dr}] dr
\end{eqnarray}
We find,
\begin{eqnarray}
\frac{d\Psi(r)}{dr}= [\frac{D-3}{2}r^{\frac{D-5}{2}}-\mu\gamma r^{\frac{D-3}{2}}-K\frac{D+1}{2}r^{\frac{D-1}{2}}+k\mu\gamma r^{\frac{D+1}{2}}]e^{-\mu\gamma r}\\
\frac{d^2\Psi(r)}{dr^2}= [\frac{(D-3)(D-5)}{2}r^{\frac{D-7}{2}}-2\mu\gamma \frac{D-3}{2}r^{\frac{D-5}{2}}-(K\frac{(D-1)(D+1)}{2}-\mu^2 \gamma^2)r^{\frac{D-3}{2}} \nonumber \\
+2K\mu\gamma \frac{D+1}{2}r^{\frac{D-1}{2}}-K\mu^2 \gamma^2r^{\frac{D+1}{2}}]e^{-\mu\gamma r}
\end{eqnarray}

\begin{eqnarray}
<H_1>=-\frac{D C_D N_1^2}{2\mu}\int_0^{\infty}[g_1r^{2D-6}-g_2r^{2D-5}- (g_3+Kg_1)r^{2D-4}\nonumber \\ +(g_4+Kg_2)r^{2D-3}+(Kg_3-g_5)r^{2D-2}-Kg_4r^{2D-1}+Kg_5r^{2D}]e^{-2\mu\gamma r}dr\\
=-\frac{D C_D N_1^2}{2\mu}[g_1I_1-g_2I_2- (g_3+Kg_1)I_3+(g_4+Kg_2)I_4+(Kg_3-g_5)I_5-Kg_4I_6+Kg_5I_7]
\end{eqnarray}
$g_i$'s are given as follows-
\begin{eqnarray}
g_1(D)=\frac{(D-1)(D-3)}{2}+\frac{(D-3)(D-5)}{4},\; g_2(D)= 2\mu\gamma (D-2) \nonumber \\
g_3(D)= \frac{3}{4}K(D+1)(D-1)-\mu^2\gamma^2,\; g_4(D)= 2D\mu\gamma K ,\; g_5(D)= K\mu^2 \gamma^2
\end{eqnarray}
Also we find,
\begin{eqnarray}
I_1= \int_0^{\infty}r^{2D-6}e^{-2\mu\gamma r}dr = \frac{\Gamma(2D-5)}{(2\mu\gamma)^{2D-5}} \\
I_2= \int_0^{\infty}r^{2D-5}e^{-2\mu\gamma r}dr = \frac{\Gamma(2D-4)}{(2\mu\gamma)^{2D-4}} \\
I_3= \int_0^{\infty}r^{2D-4}e^{-2\mu\gamma r}dr = \frac{\Gamma(2D-3)}{(2\mu\gamma)^{2D-3}} \\
I_4= \int_0^{\infty}r^{2D-3}e^{-2\mu\gamma r}dr = \frac{\Gamma(2D-2)}{(2\mu\gamma)^{2D-2}} \\
I_5= \int_0^{\infty}r^{2D-2}e^{-2\mu\gamma r}dr = \frac{\Gamma(2D-1)}{(2\mu\gamma)^{2D-1}} \\
I_6= \int_0^{\infty}r^{2D-1}e^{-2\mu\gamma r}dr = \frac{\Gamma(2D)}{(2\mu\gamma)^{2D}} \\
I_7= \int_0^{\infty}r^{2D}e^{-2\mu\gamma r}dr   = \frac{\Gamma(2D+1)}{(2\mu\gamma)^{2D+1}}
\end{eqnarray}
Appling equations (A.8-A.14) in equation (A.6) we get:
\begin{eqnarray}
<H_1>=-\frac{D C_D
N_1^2}{2\mu}[g_1\frac{\Gamma(2D-5)}{(2\mu\gamma)^{2D-5}}-g_2\frac{\Gamma(2D-4)}{(2\mu\gamma)^{2D-4}}-(g_3+Kg_1)\frac{\Gamma(2D-3)}{(2\mu\gamma)^{2D-3}}\nonumber\\
+(g_4+Kg_2)\frac{\Gamma(2D-2)}{(2\mu\gamma)^{2D-2}}+(Kg_3-g_5)\frac{\Gamma(2D-1)}{(2\mu\gamma)^{2D-1}}
-Kg_4\frac{\Gamma(2D)}{(2\mu\gamma)^{2D}}+Kg_5\frac{\Gamma(2D+1)}{(2\mu\gamma)^{2D+1}}]
\end{eqnarray}
Similarly,
\begin{eqnarray}
<H_2>= <-\frac{\gamma}{r}>= \int_0^{\infty} D C_D r^{D-1}\Psi(r)[-\frac{\gamma}{r}]\Psi(r)dr \\
= - \int_0^{\infty}\gamma D C_D r^{D-2}[\Psi(r)]^2 dr \\
=-\gamma D C_D N_1^2 \int_0^{\infty}[r^{2D-5}-2Kr^{2D-3}+K^2r^{2D-1}]e^{-2\mu\gamma r}dr  \\
=-\gamma D C_D N_1^2[\int_0^{\infty}r^{2D-5}e^{-2\mu\gamma r}dr-2K\int_0^{\infty}r^{2D-3}e^{-2\mu\gamma r}dr+K^2\int_0^{\infty}r^{2D-1}e^{-2\mu\gamma r}dr]\\
=-\gamma D C_D N_1^2 [I_2 -2KI_4 +K^2I_6]\\
=-\gamma D C_D N_1^2 [\frac{\Gamma(2D-4)}{(2\mu\gamma)^{2D-4}} -2K\frac{\Gamma(2D-2)}{(2\mu\gamma)^{2D-2}}+K^2 \frac{\Gamma(2D)}{(2\mu\gamma)^{2D}}]
\end{eqnarray}

And,
\begin{eqnarray}
<H_3> = <\sigma r + \mu_0>= \int_0^{\infty} D C_D r^{D-1}[\sigma r + \mu_0 ][\Psi(r)]^2 dr \\
=D C_D N_1^2\int_0^{\infty}[\mu_0 +\sigma r-2\mu_0 K r^2-2\sigma K r^3+\mu_0 K^2r^4+\sigma K^2r^5]r^{2D-4}e^{-2\mu\gamma r}dr \\
=DC_D N_1^2\int_0^{\infty}[\mu_0 r^{2D-4}+\sigma r^{2D-3}-2\mu_0 K r^{2D-2}-2\sigma K r^{2D-1}+\mu_0 K^2r^{2D}+\sigma K^2r^{2D+1}]e^{-2\mu\gamma r}dr\\
=DC_D N_1^2[\mu_0 I_3 +\sigma I_4-2\mu_0 K I_5 -2\sigma K I_6+\mu_0 K^2I_7+\sigma K^2I_8]\\
=DC_D N_1^2[\mu_0 \frac{\Gamma(2D-3)}{(2\mu\gamma)^{2D-3}} +\sigma \frac{\Gamma(2D-2)}{(2\mu\gamma)^{2D-2}}-2\mu_0 K \frac{\Gamma(2D-1)}{(2\mu\gamma)^{2D-1}} \nonumber \\
-2\sigma K \frac{\Gamma(2D)}{(2\mu\gamma)^{2D}}+\mu_0 K^2\frac{\Gamma(2D+1)}{(2\mu\gamma)^{2D+1}}+\sigma K^2\frac{\Gamma(2D+2)}{(2\mu\gamma)^{2D+2}}]
\end{eqnarray}

As also,
\begin{eqnarray}
I_8= \int_0^{\infty}r^{2D+1}e^{-2\mu\gamma r}dr   = \frac{\Gamma(2D+2)}{(2\mu\gamma)^{2D+2}}
\end{eqnarray}
Now, equations (A.6), (A.20) and (A.25) can be represented in more compact form as:
\begin{eqnarray}
<H_1>=-\frac{D C_D N_1^2}{2\mu}\Sigma_{i=1}^7 J_i I_i \\
<H_2>=-\gamma D C_D N_1\Sigma_{i=1}^3 J_{7+i} I_i \\
<H_3>=D C_D N_1^2\Sigma_{i=1}^6 J_{10+i} I_{2+i}
\end{eqnarray}
Here, $J_i$ terms involved in equations (A.28 - A.30) are given as follows.
\begin{eqnarray}
J_1= g_1 ,\; J_2= -g_2 ,\; J_3= -(g_3+Kg_1) ,\; J_4=  g_4+Kg_2,\; J_5=  Kg_3-g_5, \; J_6=  -Kg_4, \; J_7=  Kg_5 ,\;J_8=  1  \nonumber \\
  J_9=  -2K, \; J_{10}=  K^2, \; J_{11}=  \mu_0, \; J_{12}=  \sigma, \; J_{13}=  -2K\mu_0, \; J_{14}=  -2K\sigma, \;J_{15}=  K^2\mu_0, \; J_{16}=  K^2\sigma
\end{eqnarray}
\textbf{\emph{$g_i$ terms involved in equation (12), in the large $D$ approximation, are given below. }}
\begin{eqnarray}
\gamma \rightarrow D\frac{\pi}{24} ,\\
2\mu\gamma \rightarrow D\frac{\mu\pi}{12},\\
K  \rightarrow \frac{8\sigma}{\pi}, \\
g_1 \rightarrow D^2\frac{3}{4}  \\
g_2 \rightarrow D^2\frac{\mu \pi}{12},
g_3 \rightarrow D^2[\frac{3}{4}(\frac{8\sigma}{\pi})-(\frac{\mu\pi}{24})^2],  \\
g_4 \rightarrow D^2(\frac{8\sigma}{\pi}) \frac{\mu \pi}{12},\\
g_5 \rightarrow D^2(\frac{8\sigma}{\pi})(\frac{\mu\pi}{24})^2 , \\
L_0 \rightarrow L_c I_1  \\
where: \nonumber\\
L_c =  (\frac{2}{\frac{\pi\mu}{12}})^2[1-2(\frac{8\sigma}{\pi})(\frac{2}{\frac{\pi\mu}{12}})^2 +(\frac{8\sigma}{\pi})^2(\frac{2}{\frac{\pi\mu}{12}})^4]I_1 \\
I_i \rightarrow I_1.(\frac{2}{\frac{\pi\mu}{12}})^{i-1}\;\;(i\neq 1)
\end{eqnarray}
\textbf{\emph{$L_i$ terms involved in equations (14-16), in the large $D$ approximation, are given below. }}
\begin{eqnarray}
L_1 = \frac{g_1}{2\mu D^2} \rightarrow        \frac{1}{2\mu}\frac{3}{4} \\
L_2 = -\frac{g_2}{2\mu D^2} \rightarrow      -\frac{\mu \pi}{12}\frac{1}{2\mu} \\
L_3 = -\frac{g_3+K g_1}{2\mu D^2}\rightarrow -\frac{1}{2\mu}[\frac{3}{2}(\frac{8\sigma}{\pi}) -(\frac{\mu \pi}{24})^2] \\
L_4 = \frac{Kg_2+g_4}{2\mu D^2} \rightarrow   \frac{1}{2\mu}2(\frac{8\sigma}{\pi})(\frac{\mu \pi}{12}) \\
L_5 = \frac{Kg_3-g_5}{2\mu D^2} \rightarrow   \frac{1}{2\mu}[\frac{3}{4}(\frac{8\sigma}{\pi})^2-2(\frac{8\sigma}{\pi})(\frac{\mu \pi}{12})^2] \\
L_6 =-\frac{Kg_4}{2\mu D^2} \rightarrow      -\frac{1}{2\mu}(\frac{8\sigma}{\pi})^2(\frac{\mu \pi}{12}) \\
L_7 = \frac{Kg_5}{2\mu D^2}\rightarrow       \frac{1}{2\mu}(\frac{8\sigma}{\pi})^2(\frac{\mu \pi}{24})^2 \\
L_8 = \frac{\gamma}{D} \rightarrow           \frac{\pi}{24}  \\
L_9 = -\frac{2K\gamma}{D} \rightarrow        -2(\frac{8\sigma}{\pi})\frac{\pi}{24} \\
L_{10}= \frac{K^2\gamma}{D} \rightarrow      (\frac{8\sigma}{\pi})^2\frac{\pi}{24} \\
L_{11}=  \mu_0
\end{eqnarray}
\begin{eqnarray}
L_{12} = \sigma \\
L_{13} = -2\mu_0 K \rightarrow  -2\mu_0 \frac{8\sigma}{\pi} \\
L_{14} = -2\sigma K \rightarrow  -2\sigma \frac{8\sigma}{\pi} \\
L_{15} = \mu_0 K^2 \rightarrow  \mu_0 (\frac{8\sigma}{\pi})^2 \\
L_{16} = \sigma K^2 \rightarrow  \sigma (\frac{8\sigma}{\pi})^2
\end{eqnarray}

\numberwithin{equation}{section}
\begin{center}
\section{Appendix:B}
\end{center}
\begin{center}
\textbf{\emph{Determination of terms involved in $<H>$ in the case of Linear term as Parent. }}
\end{center}
Considering up to third term in the first infinite series of the wave function mentioned in equation (24), we obtain:
\begin{equation}
\Psi(r,D)= N_1r^{\frac{(1-D)}{2}}[1+A_1(r,D)r+A_2(r,D)r^{2}](\varrho_1 r)^{m}Ai[\varrho_1 r-\varrho_0]
\end{equation}
We simplify $A_1(r,D)$ and $A_2(r,D)$ in terms of power series in $r$ as below:
\begin{eqnarray}
A_1(r,D)= -h_1(D)r^2 + h_2(D)r^3 + h_3(D)r^4\\
A_2(r,D)= -h_4(D)r^2 + h_5(D)r^3 + h_3(D)r^4
\end{eqnarray}
where,
\begin{eqnarray}
h_1(D)=\frac{2\mu\gamma}{\varrho_1^2 k^2}\\
h_2(D)=\frac{2\mu\gamma(2g+D+1)}{\varrho_1^3 k^3}\\
h_3(D)= \frac{2\mu\gamma}{\varrho_1^4 k^4}[2g+(D-1)+2\varrho_1 +g(g-1) -(D-1)g+(2g+D-1)\varrho_1]\\
h_4(D)= \frac{2\mu W^{\prime}}{\varrho_1^2 k^2} \\
h_5(D)= \frac{2\mu W^{\prime}(2g+D+3)}{\varrho_1^3 k^3}\\
h_6(D)= \frac{2\mu W^{\prime}}{\varrho_1^4 k^4}[2+g(g-D+4)+2(D-1+2\varrho_1)+(2g+D-1)\varrho_1] \\
with\;\; g=\frac{1-D+2m}{2}
\end{eqnarray}
Other terms like $\varrho_1, \varrho_0, W^{\prime}, k \;\; etc$ are mentioned in ref. \cite{SR3}. Using equations (B.2-B.3) in (B.1) we obtain:

\begin{eqnarray}
\Psi(r,D)= N_1\varrho_1^{m}r^{g}[1 -h_1r^3 +(h_2-h_4)r^4 + (h_3+h_5)r^5+ h_6r^6]Ai[\varrho]\\
= N_1\varrho_1^{m}r^{g}[1 -h_1r^3 +h_7r^4 + h_8r^5+ h_6r^6]Ai[\varrho]\\
where\;\; h_7=h_2-h_4, h_8=h_3+h_5 \nonumber
\end{eqnarray}
The derivatives of $\Psi(r,D)$ are obtained as:
\begin{eqnarray}
\frac{d}{dr}[\Psi(r,D)]= N_1\varrho_1^{m}r^{g}[(\frac{g}{r}+Z(r))(1 -h_1r^3 +h_7r^4 + h_8r^5+ h_6r^6)\nonumber \\
+(-3h_1r^2 + 4h_7r^3 + 5h_8r^4+ 6h_6r^5)]Ai[\varrho]\\
\frac{d^2}{dr^2}[\Psi(r,D)]=N_1\varrho_1^{m}r^{g}[\frac{g^2}{r^2}(1 -h_1r^3 +h_7r^4 + h_8r^5+ h_6r^6)+\frac{2g}{r}(-3h_1 r^2+4 h_7 r^3+5 h_8 r^4+6 h_6 r^5) \nonumber \\
-\frac{g}{r^2}(1 -h_1r^3 +h_7r^4 + h_8r^5+ h_6r^6)+ (-6h_1r+12h_7r^2+20h_8r^3+30h_6r^4)\nonumber \\
+((\frac{g}{r}+1)(1 -h_1r^3 +h_7r^4 + h_8r^5+ h_6r^6)+(-3h_1 r^2+4 h_7 r^3+5 h_8 r^4+6 h_6 r^5))Z(r) \nonumber \\
+(1 -h_1r^3 +h_7r^4 + h_8r^5+ h_6r^6)(Z^2(r) + Z^{\prime}(r)) ]Ai[\varrho]
\end{eqnarray}
We use (B.13) and (B.14) to numerically calculate $<H_1>$ given in equation (26) for the case of linear parent.


\begin{thebibliography}{99}
\bibitem{Nambu-Goto}      M. L\"{u}scher and P. Weisz JHEP 07(2002)049.\\
\bibitem{Bjorken}         J. D. Bjorken, SLAC-PUB-5278(1990).
\bibitem{Neubert}         M. Neubert, Phy. Rept. 245(1994)259.
\bibitem{Vinodkumar1}     A. K. Rai, B. Patel and P. C. Vinodkumar, Phys. Rev C 78(2008)055202.
\bibitem{Vinodkumar2}     A. K. Rai, R.H. Parmar and P. C. Vinodkumar, J. Phy. G: Nucl. Part. Phy. 28(2002)2275.
\bibitem{Isgur1}          N. Isgur and M. B. Wise, Phys. Rev. Lett 66(1991)1130.
\bibitem{Bali}            G. S. Bali, arXiv:0001312[hep-ph].
\bibitem{Bali2}           G.S. Bali, Phys. Lett. B460(1999)170[hep-ph/9905387].
\bibitem{NG1}             M. L\"{u}scher, Nucl. Phy. B , 180 (1981) 317
\bibitem{NG2}             J. F. Arvis, Phy. Lett B, 127 (1983) 106.
\bibitem{NG3}             A. Antillon et al, Phy. Rev. D , 49(1994)1966.
\bibitem{Eicheten}        E. Eicheten et al, Phy. REv. D 17,3090(1978); Phy. Rev. D 21, 203(1980).
\bibitem{Luscher1}        M. L\"{u}scher, K. Symanzik and P. Weisz, Nucl. Phys. B 173(1980)365.
\bibitem{Sigma}           A. Yamamato, H. Suganuma and H. Iida Phys. Lett. B 664(2008)129.
\bibitem{Dong}            S. H. Dong \emph{et al}, Found. Phys. Lett. 12(1999)465.
\bibitem{Gang}            C. Gang ,Chin. Phys. 14(2005)1075.
\bibitem{SR1}             S. Roy, B. J. Hazarika and D. K. Choudhury, Phys. Scr. 86(2012)045101 (arXiv:1205.5330[hep-ph]).
\bibitem{SR2}             S. Roy and D. K. Choudhury, arXiv: 1305.1412[hep-ph].
\bibitem{SR3}             S. Roy and D. K. Choudhury, Phys. Scr. 87(2013)065101 (arXiv:1301.0982[hep-ph]).
\bibitem{KKP}             K. K. Pathak and D. K. Choudhury, Pramana J. Phys. 79(6)(2012)1385.
\bibitem{GRKhan}          G. R. Khan, Eur. Phys. J D 53 (2009)123.
\bibitem{Laplace}     Steven Rosenberg ``The Laplacian on a Riemannian Manifold: An Introduction to Analysis on Manifolds",  Cambridge University Press (1997).
\bibitem{Noura}           Moura Eduarda, David G. Henderson in "Experiencing Geometry:On plane and sphere",Prentice Hall(1996).
\bibitem{PDG}             J. Beringer \emph{et al} (Particle Data Group), Phy. Rev. D86(2012)010001(http://pdg.ibi.gov).
\bibitem{Morales}         Daniel A. Morales, Int. J. Quant. Chem 57(1996)7.
\bibitem{Burgbacher}      F. Burgbacher \emph{et. al} J. Math. Phys. 40(1999)625.
\bibitem{Caruso}          F. Caruso \emph{et. al} Phys. Letts. A 377 (2013) 694.
\bibitem{Arvis}         J. F. Arvis, Phys. Lett. B 127 (1983)106. \\


\end{thebibliography}
\end{document}